\documentclass{article}

\usepackage{arxiv}

\usepackage[utf8]{inputenc} % allow utf-8 input
\usepackage[T1]{fontenc}    % use 8-bit T1 fonts
\usepackage{hyperref}       % hyperlinks
\usepackage{url}            % simple URL typesetting
\usepackage{booktabs}       % professional-quality tables
\usepackage{amsfonts}       % blackboard math symbols
\usepackage{amssymb}        % needed for \leqslant
\usepackage{amsmath}        % needed for cases
\usepackage{nicefrac}       % compact symbols for 1/2, etc.
\usepackage{microtype}      % microtypography
\usepackage{lipsum}		% Can be removed after putting your text content
\usepackage{graphicx}
\usepackage[numbers]{natbib}
\usepackage{doi}
\usepackage{wrapfig}

\title{Jordan algebra in R}

\author{ \href{https://orcid.org/0000-0001-5982-0415}{\includegraphics[width=0.03\textwidth]{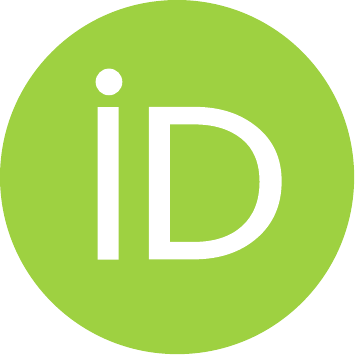}\hspace{1mm}Robin K. S.~Hankin}\thanks{\href{https://academics.aut.ac.nz/robin.hankin}{work};  
\href{https://www.youtube.com/watch?v=JzCX3FqDIOc&list=PL9_n3Tqzq9iWtgD8POJFdnVUCZ_zw6OiB&ab_channel=TrinTragulaGeneralRelativity}{play}} \\
 Auckland University of Technology\\
	\texttt{hankin.robin@gmail.com} \\
}

\renewcommand{\shorttitle}{Jordan algebra in R}

\hypersetup{
pdftitle={Jordan algebra in R},
pdfsubject={q-bio.NC, q-bio.QM},
pdfauthor={Robin K. S.~Hankin},
pdfkeywords={Jordan algebra}
}

\usepackage{Sweave}
\begin{document}
\maketitle

\begin{abstract} In this short article I introduce the {\tt jordan}
package which provides functionality for working with different types
of Jordan algebra.  I give some numerical verification of the Jordan
identity for the five types of Jordan algebras.  The package is
available on CRAN at \url{https://CRAN.R-project.org/package=stokes}.

\end{abstract}

\section{Introduction: Jordan algebras}

\setlength{\intextsep}{0pt}
\begin{wrapfigure}{r}{0.2\textwidth}
  \begin{center}
\includegraphics[width=1in]{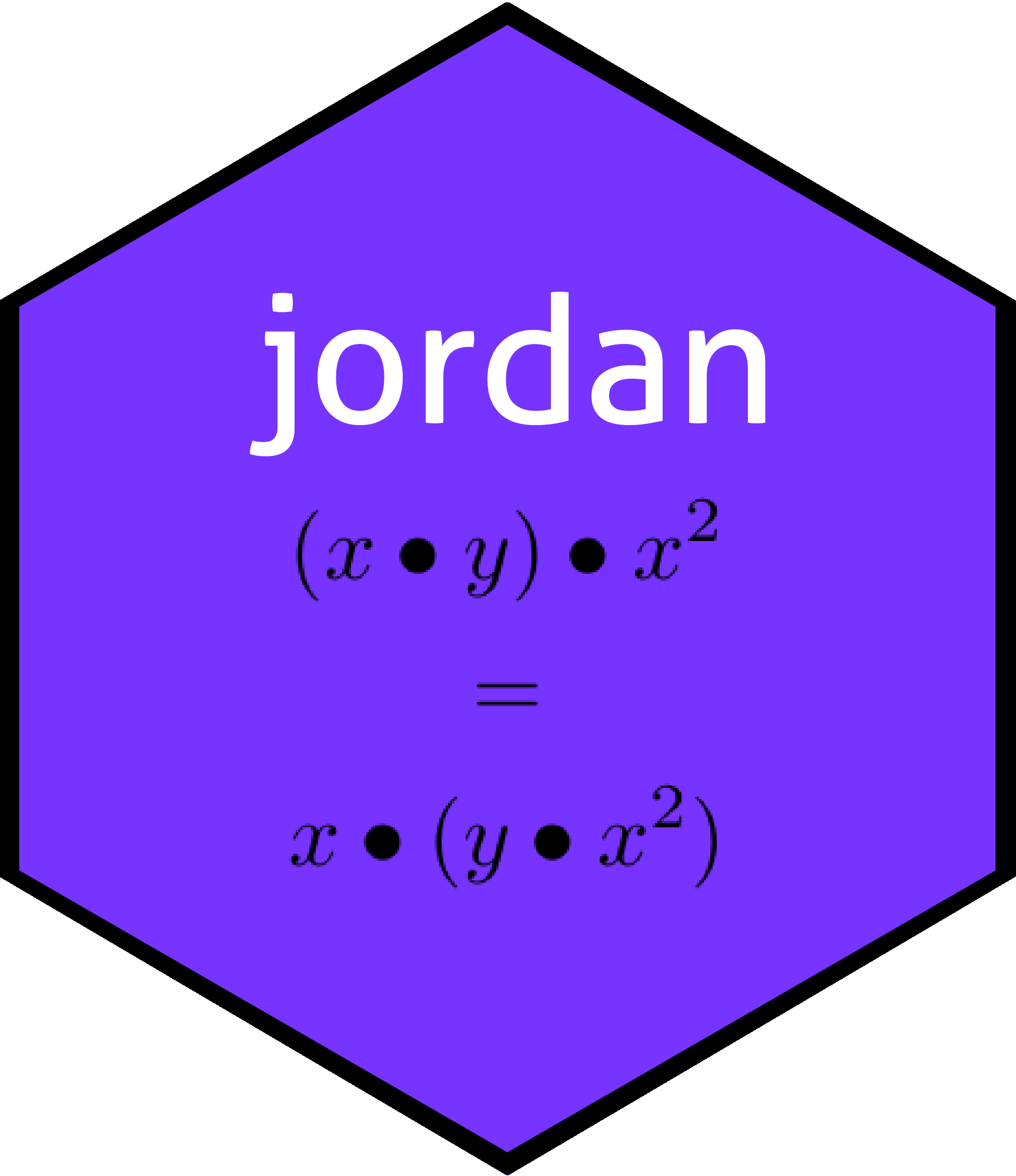}
  \end{center}
\end{wrapfigure}

A {\em Jordan algebra} is a non-associative algebra over the reals with
a bilinear multiplication that satisfies the following identities:

$$xy=yx$$

$$(xy)(xx)=x(y(xx))$$

(the second identity is known as the Jordan identity).  In literature,
multiplication is usually indicated by juxtaposition but one sometimes
sees $x\circ y$.  Package idiom is to use an asterisk, as in {\tt x*y}.
Following McCrimmon~\cite{mccrimmon1978}, there are five types of
Jordan algebras:

\begin{itemize}
\item {\bf type 1}: Real symmetric matrices, class {\tt real\_symmetric\_matrix},
    abbreviated in the package to {\tt rsm}
\item {\bf type 2}: Complex Hermitian matrices, class {\tt complex\_herm\_matrix},
    abbreviated to {\tt chm}
\item {\bf type 3}: Quaternionic Hermitian matrices, class
    {\tt quaternion\_herm\_matrix}, abbreviated to {\tt qhm}
\item {\bf type 4}: Albert algebras, the space of $3\times 3$
    Hermitian octonionic matrices, class {\tt albert}
\item {\bf type 5}: Spin factors, class {\tt spin}
\end{itemize}

(of course, the first two are special cases of the next).  The
{\tt  jordan} package provides functionality to manipulate jordan objects
using natural R idiom.  Objects of all these classes are stored in
matrix form with columns being elements of the jordan algebra.  The
first four classes are matrix-based in the sense that the algebraic
objects are symmetric or Hermitian matrices (the S4 class is
{\tt jordan\_matrix}).  The fifth class, spin factors, is not matrix based.

\section{Matrix-based Jordan algebras, types 1,2,3}

The four matrix-based Jordan algebras have elements which are square
matrices.  The first three classes are real (symmetric) matrices,
complex (Hermitian) matrices, and quaternionic (Hermitian); type 4 is
considered separately at the end.  Types 1,2, and 3 all behave in the
same way from a package idiom perspective.  Consider:

\begin{Schunk}
\begin{Sinput}
> library(jordan)
> x <- rrsm()  # "Random Real Symmetric Matrix"
> y <- rrsm()  
> z <- rrsm()  
> x
\end{Sinput}
\begin{Soutput}
Vector of real symmetric matrices with entries
        [1]   [2]   [3] 
 [1,] -1.41  1.76 -0.01 
 [2,]  1.70 -1.67 -0.41 
 [3,]  0.57  0.28 -0.75 
 [4,] -0.34 -0.21 -0.23 
 [5,]  0.16  1.18  1.63 
....................... 
[11,]  1.03 -0.02  1.81 
[12,]  1.78 -0.43  2.14 
[13,]  0.31 -0.42  1.12 
[14,]  0.01 -0.27 -0.56 
[15,]  0.69  1.46  0.21 
\end{Soutput}
\end{Schunk}

Object {\tt x} is a three-element vector, with each element being a
column.  Each element corresponds to a $5\times 5$ symmetric matrix
(because {\tt rrsm()} has {\tt d=5} by default, specifying the size of
the matrix).  Thus each element has $5*(5+1)/2=15$ degrees of freedom,
as indicated by the row labelling.  Addition and multiplication of a
Jordan object with a scalar are defined:

\begin{Schunk}
\begin{Sinput}
> x*100
\end{Sinput}
\begin{Soutput}
Vector of real symmetric matrices with entries
       [1]  [2]  [3] 
 [1,] -141  176   -1 
 [2,]  170 -167  -41 
 [3,]   57   28  -75 
 [4,]  -34  -21  -23 
 [5,]   16  118  163 
.................... 
[11,]  103   -2  181 
[12,]  178  -43  214 
[13,]   31  -42  112 
[14,]    1  -27  -56 
[15,]   69  146   21 
\end{Soutput}
\begin{Sinput}
> x + y*3
\end{Sinput}
\begin{Soutput}
Vector of real symmetric matrices with entries
        [1]   [2]   [3] 
 [1,] -0.24 -2.65 -0.97 
 [2,] -1.24 -4.19  0.61 
 [3,]  2.82 -3.05 -0.66 
 [4,]  0.62  0.96  5.08 
 [5,]  0.37 -2.09  4.99 
....................... 
[11,]  5.56 -2.60  1.48 
[12,]  4.75 -3.73  8.74 
[13,]  0.37  0.54 -1.37 
[14,]  3.82 -0.81  3.07 
[15,]  2.01 -0.76  2.55 
\end{Soutput}
\begin{Sinput}
> x + 100
\end{Sinput}
\begin{Soutput}
Vector of real symmetric matrices with entries
         [1]    [2]    [3] 
 [1,]  98.59 101.76  99.99 
 [2,] 101.70  98.33  99.59 
 [3,] 100.57 100.28  99.25 
 [4,]  99.66  99.79  99.77 
 [5,] 100.16 101.18 101.63 
.......................... 
[11,] 101.03  99.98 101.81 
[12,] 101.78  99.57 102.14 
[13,] 100.31  99.58 101.12 
[14,] 100.01  99.73  99.44 
[15,] 100.69 101.46 100.21 
\end{Soutput}
\end{Schunk}

(the last line is motivated by analogy with {\tt M + x}, for {\tt M} a matrix
and {\tt x} a scalar). Jordan objects may be multiplied using the rule
$x\circ y=(xy+yx)/2$:

\begin{Schunk}
\begin{Sinput}
> x*y
\end{Sinput}
\begin{Soutput}
Vector of real symmetric matrices with entries
           [1]      [2]      [3] 
 [1,] -0.39760 -0.47830 -1.42100 
 [2,]  2.44290  0.55960  3.98550 
 [3,]  2.09730 -1.98220  5.26770 
 [4,] -0.45340  0.02920 -0.32840 
 [5,]  1.07915 -0.68270 -1.66305 
................................ 
[11,] -1.17490 -0.58995  0.53450 
[12,]  4.05200  0.73435  1.61260 
[13,] -0.40450  0.16130  4.87155 
[14,]  1.21355 -1.21490  0.46040 
[15,]  3.64000 -0.67600  3.06550 
\end{Soutput}
\end{Schunk}

We may verify that the distributive law is obeyed:

\begin{Schunk}
\begin{Sinput}
> x*(y+z) - (x*y + x*z)
\end{Sinput}
\begin{Soutput}
Vector of real symmetric matrices with entries
                [1]           [2]           [3] 
 [1,] -6.661338e-16  2.081668e-16  0.000000e+00 
 [2,]  0.000000e+00 -4.440892e-16  0.000000e+00 
 [3,]  0.000000e+00  0.000000e+00 -1.776357e-15 
 [4,] -3.330669e-16  0.000000e+00  1.110223e-16 
 [5,] -8.881784e-16  0.000000e+00 -4.440892e-16 
............................................... 
[11,]  4.440892e-16  1.110223e-16  8.881784e-16 
[12,]  0.000000e+00  0.000000e+00  0.000000e+00 
[13,]  0.000000e+00 -2.220446e-16 -8.881784e-16 
[14,] -4.440892e-16 -2.220446e-16  0.000000e+00 
[15,]  0.000000e+00  0.000000e+00 -1.776357e-15 
\end{Soutput}
\end{Schunk}

(that is, zero to numerical precision).  Further, we may observe that
the resulting algebra is not associative:

\begin{Schunk}
\begin{Sinput}
> LHS <- x*(y*z)
> RHS <- (x*y)*z
> LHS-RHS
\end{Sinput}
\begin{Soutput}
Vector of real symmetric matrices with entries
              [1]        [2]        [3] 
 [1,]  -3.9564130  0.8402355 -2.4048880 
 [2,]   3.9440350 -2.0422630 -4.1022730 
 [3,]   2.2380955  1.0996705 -8.9562500 
 [4,]   1.4684547  0.8016225  1.1587312 
 [5,]  -1.4314930  0.6700580  4.4210495 
....................................... 
[11,]   4.1317442  1.0128450 -1.9271358 
[12,]   0.5929205  0.2259248 -1.4115125 
[13,]   0.8094577  0.7190828  6.6453095 
[14,]  -7.9400165  0.0537760 -0.7245633 
[15,]   6.0556225  1.1611970  9.5536030 
\end{Soutput}
\end{Schunk}

showing numerically that $x(yz)\neq(xy)z$.  However, the Jordan
identity $(xy)(xx) = x(y(xx))$ is satisfied:

\begin{Schunk}
\begin{Sinput}
> LHS <- (x*y)*(x*x)
> RHS <- x*(y*(x*x))
> LHS-RHS
\end{Sinput}
\begin{Soutput}
Vector of real symmetric matrices with entries
                [1]           [2]           [3] 
 [1,] -1.776357e-15 -4.440892e-15 -8.881784e-16 
 [2,]  0.000000e+00  1.776357e-15 -3.552714e-15 
 [3,]  7.105427e-15 -1.776357e-15  0.000000e+00 
 [4,]  0.000000e+00 -2.220446e-16 -1.332268e-15 
 [5,] -1.776357e-15 -8.881784e-16 -3.552714e-15 
............................................... 
[11,]  4.440892e-16  0.000000e+00 -1.776357e-15 
[12,]  7.105427e-15 -8.881784e-16  0.000000e+00 
[13,] -1.332268e-15  6.661338e-16  1.421085e-14 
[14,]  1.776357e-15 -1.776357e-15 -1.776357e-15 
[15,]  0.000000e+00  2.220446e-16  7.105427e-15 
\end{Soutput}
\end{Schunk}

(the entries are zero to numerical precision).  If we wish to work
with the matrix itself, a single element may be coerced with
{\tt as.1matrix()}:

\begin{Schunk}
\begin{Sinput}
> M1 <- as.1matrix(x[1])
> (M2 <- as.1matrix(x[2]))
\end{Sinput}
\begin{Soutput}
      [,1]  [,2]  [,3]  [,4]  [,5]
[1,]  1.76 -1.67 -0.21  1.64 -0.02
[2,] -1.67  0.28  1.18  1.19 -0.43
[3,] -0.21  1.18 -1.36 -0.46 -0.42
[4,]  1.64  1.19 -0.46  1.57 -0.27
[5,] -0.02 -0.43 -0.42 -0.27  1.46
\end{Soutput}
\end{Schunk}

(in the above, observe how the matrix is indeed symmetric).  We may
verify that the multiplication rule is indeed being correctly
applied:

\begin{Schunk}
\begin{Sinput}
> (M1 %*% M2 + M2 %*% M1)/2 - as.1matrix(x[1]*x[2])
\end{Sinput}
\begin{Soutput}
     [,1] [,2] [,3] [,4] [,5]
[1,]    0    0    0    0    0
[2,]    0    0    0    0    0
[3,]    0    0    0    0    0
[4,]    0    0    0    0    0
[5,]    0    0    0    0    0
\end{Soutput}
\end{Schunk}

It is also possible to verify that symmetry is preserved under the
Jordan operation:

\begin{Schunk}
\begin{Sinput}
> jj <- as.1matrix(x[1]*x[2])
> jj-t(jj)
\end{Sinput}
\begin{Soutput}
     [,1] [,2] [,3] [,4] [,5]
[1,]    0    0    0    0    0
[2,]    0    0    0    0    0
[3,]    0    0    0    0    0
[4,]    0    0    0    0    0
[5,]    0    0    0    0    0
\end{Soutput}
\end{Schunk}

The other matrix-based Jordan algebras are similar but
differ in the details of the coercion.  Taking quaternionic matrices:

\begin{Schunk}
\begin{Sinput}
> as.1matrix(rqhm(n=1,d=2))
\end{Sinput}
\begin{Soutput}
   [1,1] [2,1] [1,2] [2,2]
Re  1.26  0.81  0.81  1.53
i   0.00  1.50 -1.50  0.00
j   0.00  0.08 -0.08  0.00
k   0.00 -1.05  1.05  0.00
     [,1] [,2]
[1,]    1    3
[2,]    2    4
\end{Soutput}
\end{Schunk}

above we see the matrix functionality of the {\tt onion} package being
used.  See how the matrix is Hermitian (elements {\tt [1,2]} and {\tt
[2,1]} are conjugate; elements {\tt [1,1]} and {\tt [2,2]} are pure
real).  Verifying the Jordan identity would be almost the same as
above:

\begin{Schunk}
\begin{Sinput}
> x <- rqhm()
> y <- rqhm()
> (x*y)*(x*x) - x*(y*(x*x))
\end{Sinput}
\begin{Soutput}
Vector of quaternionic Hermitian matrices with entries
                [1]           [2]           [3] 
 [1,]  0.000000e+00 -1.421085e-14 -1.421085e-14 
 [2,]  0.000000e+00  0.000000e+00  0.000000e+00 
 [3,]  7.105427e-15  5.329071e-15  1.421085e-14 
 [4,]  1.421085e-14  0.000000e+00 -1.421085e-14 
 [5,]  7.105427e-15  2.842171e-14 -3.552714e-15 
............................................... 
[41,] -3.552714e-15 -7.105427e-15  7.105427e-15 
[42,]  7.105427e-15 -1.421085e-14 -3.552714e-15 
[43,] -3.552714e-15 -7.105427e-15  1.776357e-15 
[44,] -1.776357e-15  0.000000e+00  7.105427e-15 
[45,]  0.000000e+00  0.000000e+00  8.881784e-15 
\end{Soutput}
\end{Schunk}

Above, we see that the difference is very small.

\section{Spin factors, type 5}

The first four types of Jordan algebras are all matrix-based with
either real, complex, quaternionic, or octonionic entries.

The fifth type, spin factors, are slightly different from a package
idiom perspective.  Mathematically, elements are of the form
$\mathbb{R}\oplus\mathbb{R}^n$, with addition and multiplication
defined by

$$\alpha(a,\mathbf{a}) = (\alpha a,\alpha\mathbf{a})$$

$$(a,\mathbf{a}) + (b,\mathbf{b}) = (a+b,\mathbf{a} + \mathbf{b})$$

$$(a,\mathbf{a}) \times (b,\mathbf{b}) =
(ab+\left\langle\mathbf{a},\mathbf{b}\right\rangle,a\mathbf{b} +
b\mathbf{a})$$

where $a,b,\alpha\in\mathbb{R}$, and
$\mathbf{a},\mathbf{b}\in\mathbb{R}^n$.  Here
$\left\langle\cdot,\cdot\right\rangle$ is an inner product defined on
$\mathbb{R}^n$ (by default we have $\left\langle(x_1,\ldots,
x_n),(y_1,\ldots, y_n)\right\rangle=\sum x_iy_i$ but this is
configurable in the package).

So if we have $\mathcal{I},\mathcal{J},\mathcal{K}$ spin factor
elements it is clear that
$\mathcal{I}\mathcal{J}=\mathcal{J}\mathcal{I}$ and
$\mathcal{I}(\mathcal{J}+\mathcal{K}) = \mathcal{I}\mathcal{J} +
\mathcal{I}\mathcal{K}$.  The Jordan identity is not as easy to see
but we may verify all the identities numerically:

\begin{Schunk}
\begin{Sinput}
> I <- rspin()
> J <- rspin()
> K <- rspin()
> I
\end{Sinput}
\begin{Soutput}
Vector of spin objects with entries
      [1]   [2]   [3] 
 r   0.49  0.38 -2.13 
[1]  0.41 -0.48 -0.99 
[2] -0.21 -0.86 -0.62 
[3] -0.25 -0.41 -1.01 
[4] -2.48  1.86 -0.65 
[5] -0.02 -0.10 -1.17 
\end{Soutput}
\begin{Sinput}
> I*J - J*I   # commutative:
\end{Sinput}
\begin{Soutput}
Vector of spin objects with entries
    [1] [2] [3] 
 r    0   0   0 
[1]   0   0   0 
[2]   0   0   0 
[3]   0   0   0 
[4]   0   0   0 
[5]   0   0   0 
\end{Soutput}
\begin{Sinput}
> I*(J+K) - (I*J + I*K)  # distributive:
\end{Sinput}
\begin{Soutput}
Vector of spin objects with entries
              [1]           [2]           [3] 
 r   2.220446e-16  3.330669e-16  0.000000e+00 
[1]  1.110223e-16  0.000000e+00  1.110223e-16 
[2]  0.000000e+00  2.220446e-16  0.000000e+00 
[3]  0.000000e+00 -1.110223e-16  0.000000e+00 
[4]  0.000000e+00  0.000000e+00  0.000000e+00 
[5] -1.110223e-16  0.000000e+00 -2.220446e-16 
\end{Soutput}
\begin{Sinput}
> I*(J*K) - (I*J)*K  # not associative:
\end{Sinput}
\begin{Soutput}
Vector of spin objects with entries
              [1]       [2]       [3] 
 r   8.881784e-16  0.000000  0.000000 
[1] -2.814170e-01 -3.279000 -1.032597 
[2]  1.222649e+00 -0.163754 -5.508214 
[3] -1.216666e+00 -0.511667  2.281628 
[4] -3.774480e+00  1.286646  1.430624 
[5]  1.619257e+00 -0.956662 -3.466692 
\end{Soutput}
\begin{Sinput}
> (I*J)*(I*I) - I*(J*(I*I))  # Obeys the Jordan identity
\end{Sinput}
\begin{Soutput}
Vector of spin objects with entries
              [1]           [2]          [3] 
 r  -3.552714e-15 -8.881784e-16 0.000000e+00 
[1]  0.000000e+00  0.000000e+00 8.881784e-16 
[2]  0.000000e+00  0.000000e+00 0.000000e+00 
[3] -1.110223e-16  0.000000e+00 0.000000e+00 
[4]  0.000000e+00  8.881784e-16 3.552714e-15 
[5] -2.220446e-16 -2.220446e-16 0.000000e+00 
\end{Soutput}
\end{Schunk}

\section{Albert algebras, type 4}

Type 4 Jordan algebra corresponds to $3\times 3$ Hermitian matrices
with octonions for entries.  This is class {\tt albert} in the package.
Note that octonionic Hermitian matrices of order 4 or above do not
satisfy the Jordan identity and are therefore not Jordan algebras:
there is a numerical illustration of this fact in the {\tt onion} package
vignette.  We may verify the Jordan identity for $3\times 3$
octonionic Hermitian matrices using package class {\tt albert}:

\begin{Schunk}
\begin{Sinput}
> x <- ralbert()
> y <- ralbert()
> x
\end{Sinput}
\begin{Soutput}
Vector of Albert matrices with entries
         [1]   [2]   [3]
    d1  1.54  0.03 -0.82
    d2 -1.48 -0.65 -0.74
    d3 -0.19  0.31  0.24
Re(o1) -0.40  0.39  0.34
 i(o1)  0.05  0.84  0.33
 j(o1) -1.31 -0.11 -0.15
 k(o1) -2.09 -1.05  0.08
 l(o1) -0.60  1.28 -1.46
il(o1) -0.94  1.69 -1.82
jl(o1)  1.30 -0.49 -2.73
kl(o1)  2.42 -0.65  0.50
Re(o2) -1.08 -0.40  0.53
 i(o2) -1.41  0.87  0.57
 j(o2)  0.84 -0.60 -0.58
 k(o2) -0.75  0.47  0.95
 l(o2) -0.62  0.84 -0.67
il(o2) -1.18  0.46  0.36
jl(o2) -0.15  0.97  0.94
kl(o2) -0.69 -2.45 -0.81
Re(o3)  0.78 -0.58  0.25
 i(o3) -1.12 -0.33 -1.70
 j(o3) -0.29  0.35  0.16
 k(o3)  1.68  0.18 -0.06
 l(o3)  0.92  0.84  0.62
il(o3) -0.25  0.01  0.05
jl(o3) -0.43  0.68  2.27
kl(o3) -0.69  2.60 -1.46
\end{Soutput}
\begin{Sinput}
> (x*y)*(x*x)-x*(y*(x*x)) # Jordan identity:
\end{Sinput}
\begin{Soutput}
Vector of Albert matrices with entries
                 [1]           [2]           [3]
    d1  1.509903e-14  7.105427e-15  8.881784e-15
    d2 -2.842171e-14  1.421085e-14 -7.105427e-15
    d3 -7.105427e-15  1.776357e-15 -7.105427e-15
Re(o1)  1.509903e-14 -1.065814e-14  0.000000e+00
 i(o1)  7.105427e-15  3.552714e-15 -5.329071e-15
 j(o1)  7.105427e-15 -1.776357e-15  0.000000e+00
 k(o1)  0.000000e+00  1.421085e-14  0.000000e+00
 l(o1)  7.105427e-15  0.000000e+00  0.000000e+00
il(o1)  0.000000e+00 -1.421085e-14 -3.552714e-15
jl(o1)  0.000000e+00  1.421085e-14  0.000000e+00
kl(o1)  2.842171e-14 -8.881784e-16  0.000000e+00
Re(o2) -2.131628e-14 -7.105427e-15  0.000000e+00
 i(o2) -7.105427e-15  1.776357e-15 -2.220446e-15
 j(o2)  3.552714e-15  3.552714e-15 -7.105427e-15
 k(o2)  0.000000e+00  7.105427e-15 -7.105427e-15
 l(o2)  1.421085e-14 -7.105427e-15 -7.105427e-15
il(o2) -1.421085e-14  1.421085e-14 -6.217249e-15
jl(o2) -7.105427e-15 -3.552714e-15  3.552714e-15
kl(o2) -1.421085e-14 -1.065814e-14  3.552714e-15
Re(o3)  7.105427e-15  0.000000e+00 -3.552714e-15
 i(o3) -2.842171e-14  5.329071e-15  0.000000e+00
 j(o3)  0.000000e+00  4.440892e-16  0.000000e+00
 k(o3)  0.000000e+00 -7.105427e-15 -3.552714e-15
 l(o3)  0.000000e+00 -1.421085e-14  0.000000e+00
il(o3)  1.421085e-14  1.065814e-14  8.881784e-15
jl(o3) -2.842171e-14  1.065814e-14  0.000000e+00
kl(o3) -1.421085e-14  7.105427e-15 -7.105427e-15
\end{Soutput}
\end{Schunk}

\section{Special identities}

In 1963, C. M. Glennie discovered a pair of identities satisfied by
special Jordan algebras but not the Albert algebra.  Defining

\[
U_x(y) = 2x(xy)-(xx)y
\]

\[
\left\lbrace x,y,z\right\rbrace=
2(x(yz)+(xy)z - (xz)y)
\]
 
(it can be shown that Jacobson's identity $U_{U_x(y)}=U_xU_yU_x$
holds), Glennie's identities are

\[
H_8(x,y,z)=H_8(y,x,z)\qquad H_9(x,y,z)=H_9(y,x,z)
\]

(see McCrimmon 2004 for details), where

\[
H_8(x,y,z)= \left\lbrace U_x U_y(z),z, xy\right\rbrace-U_xU_yU_z(xy)
\]

and
\[
H_9(x,y,z)= 2U_x(z) U_{y,x}U_z(yy)-U_x U_z U_{x,y} U_y(z)
\]

\subsection{Numerical verification of Jacobson}

We may verify Jacobson's identity:

\begin{Schunk}
\begin{Sinput}
> U <- function(x){function(y){2*x*(x*y)-(x*x)*y}}
> diff <- function(x,y,z){
+      LHS <- U(x)(U(y)(U(x)(z)))
+      RHS <- U(U(x)(y))(z)
+      return(LHS-RHS)  # zero if Jacobson holds
+ }
\end{Sinput}
\end{Schunk}

Then we may numerically verify Jacobson for type 3-5 Jordan algebras:

\begin{Schunk}
\begin{Sinput}
> diff(ralbert(),ralbert(),ralbert())  # Albert algebra obeys Jacobson:
\end{Sinput}
\begin{Soutput}
Vector of Albert matrices with entries
                 [1]           [2]           [3]
    d1  3.637979e-12 -1.477929e-12  2.273737e-12
    d2 -3.637979e-12 -9.094947e-13  4.547474e-13
    d3  1.364242e-12  6.821210e-13  0.000000e+00
Re(o1) -1.364242e-12 -3.979039e-13  6.821210e-13
 i(o1)  0.000000e+00 -1.136868e-13  0.000000e+00
 j(o1)  4.547474e-13  4.547474e-13 -6.821210e-13
 k(o1) -2.273737e-13  1.364242e-12 -4.547474e-13
 l(o1) -2.273737e-12 -1.364242e-12  4.547474e-13
il(o1)  0.000000e+00 -1.364242e-12  9.094947e-13
jl(o1)  1.818989e-12  3.410605e-13 -4.547474e-13
kl(o1) -2.728484e-12 -2.273737e-13  1.364242e-12
Re(o2) -1.364242e-12  1.421085e-12  1.818989e-12
 i(o2)  9.094947e-13  1.136868e-13  9.094947e-13
 j(o2)  1.818989e-12 -7.958079e-13 -9.094947e-13
 k(o2)  9.094947e-13 -3.410605e-13  0.000000e+00
 l(o2) -9.094947e-13 -1.364242e-12  1.136868e-12
il(o2) -1.818989e-12  0.000000e+00  9.094947e-13
jl(o2) -9.094947e-13  1.477929e-12 -9.094947e-13
kl(o2)  4.547474e-13 -3.410605e-13  0.000000e+00
Re(o3)  0.000000e+00  1.136868e-12 -2.273737e-13
 i(o3)  1.818989e-12 -2.501110e-12 -6.821210e-13
 j(o3) -1.818989e-12 -7.673862e-13  4.547474e-13
 k(o3) -1.364242e-12 -2.273737e-13 -4.547474e-13
 l(o3) -4.547474e-13  0.000000e+00  0.000000e+00
il(o3)  1.818989e-12 -2.273737e-13  4.547474e-13
jl(o3)  4.547474e-13  3.183231e-12  0.000000e+00
kl(o3)  1.818989e-12  1.136868e-12  4.547474e-13
\end{Soutput}
\begin{Sinput}
> diff(rqhm(),rqhm(),rqhm()) # Quaternion Jordan algebra obeys Jacobson:
\end{Sinput}
\begin{Soutput}
Vector of quaternionic Hermitian matrices with entries
                [1]           [2]           [3] 
 [1,] -9.094947e-13  1.364242e-12  1.818989e-12 
 [2,] -2.728484e-12 -1.193712e-12 -9.094947e-13 
 [3,]  5.456968e-12  0.000000e+00 -2.273737e-13 
 [4,]  0.000000e+00 -5.911716e-12 -9.094947e-13 
 [5,]  0.000000e+00 -4.547474e-13  1.818989e-12 
............................................... 
[41,] -1.136868e-12  0.000000e+00  0.000000e+00 
[42,] -2.955858e-12  0.000000e+00 -2.160050e-12 
[43,] -2.273737e-13 -4.547474e-13  2.728484e-12 
[44,]  2.273737e-13 -2.614797e-12 -2.728484e-12 
[45,] -1.136868e-12  1.818989e-12  1.364242e-12 
\end{Soutput}
\begin{Sinput}
> diff(rspin(),rspin(),rspin()) # spin factors obey Jacobson:
\end{Sinput}
\begin{Soutput}
Vector of spin objects with entries
              [1]           [2]          [3] 
 r   0.000000e+00 -2.842171e-14 1.421085e-14 
[1]  4.263256e-14 -7.105427e-14 8.881784e-15 
[2] -3.552714e-14 -2.131628e-14 7.105427e-15 
[3] -4.263256e-14  5.684342e-14 4.884981e-15 
[4]  2.842171e-14  0.000000e+00 5.329071e-15 
[5] -4.973799e-14 -2.842171e-14 0.000000e+00 
\end{Soutput}
\end{Schunk}

showing agreement to numerical accuracy (the output is close to zero).
We can now verify Glennie's $G_8$ and $G_9$ identities.

\subsection{Numerical verification of $G_8$}

\begin{Schunk}
\begin{Sinput}
> B <- function(x,y,z){2*(x*(y*z) + (x*y)*z - (x*z)*y)}  # bracket function
> H8 <- function(x,y,z){B(U(x)(U(y)(z)),z,x*y) - U(x)(U(y)(U(z)(x*y)))}
> G8 <- function(x,y,z){H8(x,y,z)-H8(y,x,z)}
\end{Sinput}
\end{Schunk}

and so we verify for type 3 and type 5 Jordans:

\begin{Schunk}
\begin{Sinput}
> G8(rqhm(1),rqhm(1),rqhm(1))   # Quaternion Jordan algebra obeys G8:
\end{Sinput}
\begin{Soutput}
Vector of quaternionic Hermitian matrices with entries
                [1] 
 [1,]  0.000000e+00 
 [2,] -4.547474e-12 
 [3,]  3.637979e-12 
 [4,] -1.818989e-12 
 [5,] -5.456968e-12 
................... 
[41,]  0.000000e+00 
[42,]  3.637979e-12 
[43,]  8.185452e-12 
[44,] -4.547474e-12 
[45,] -6.366463e-12 
\end{Soutput}
\begin{Sinput}
> G8(rspin(1),rspin(1),rspin(1)) # Spin factors obey G8:
\end{Sinput}
\begin{Soutput}
Vector of spin objects with entries
                a 
 r   3.552714e-15 
[1]  7.494005e-16 
[2] -4.440892e-16 
[3] -2.220446e-16 
[4] -2.664535e-15 
[5]  0.000000e+00 
\end{Soutput}
\end{Schunk}

again showing acceptable accuracy.  The identity is {\em not} true for
Albert algebras:

\begin{Schunk}
\begin{Sinput}
> G8(ralbert(1),ralbert(1),ralbert(1)) # Albert algebra does not obey G8:
\end{Sinput}
\begin{Soutput}
Vector of Albert matrices with entries
               [1]
    d1 -3209.77325
    d2 -2183.08560
    d3  5392.85884
Re(o1)  3048.77534
 i(o1)  -341.17026
 j(o1) -5069.00757
 k(o1) -3255.20454
 l(o1)  6325.31177
il(o1)  4202.32334
jl(o1) -3395.24245
kl(o1)  3494.94665
Re(o2) -1279.20890
 i(o2) -9545.59707
 j(o2)  -998.40758
 k(o2)  2071.51608
 l(o2)  1031.05238
il(o2) -7691.31254
jl(o2)  4705.92146
kl(o2)   685.36332
Re(o3) -7504.55626
 i(o3)  4532.20967
 j(o3) -5012.57428
 k(o3)   902.05461
 l(o3)  -184.23557
il(o3)    46.94533
jl(o3)  8277.70343
kl(o3)   988.06732
\end{Soutput}
\end{Schunk}

\subsection{Numerical verification of $G_9$}

\begin{Schunk}
\begin{Sinput}
> L <- function(x){function(y){x*y}}
> U <- function(x){function(y){2*x*(x*y)-(x*x)*y}}
> U2 <- function(x,y){function(z){L(x)(L(y)(z)) + L(y)(L(x)(z)) - L(x*y)(z)}}
> H9 <- function(x,y,z){2*U(x)(z)*U2(y,x)(U(z)(y*y)) - U(x)(U(z)(U2(x,y)(U(y)(z))))}
> G9 <- function(x,y,z){H9(x,y,z)-H9(y,x,z)}
\end{Sinput}
\end{Schunk}

Then we may verify the `G9()` identity for type 3 Jordans:

\begin{Schunk}
\begin{Sinput}
> G9(rqhm(1),rqhm(1),rqhm(1))  # Quaternion Jordan algebra obeys G9:
\end{Sinput}
\begin{Soutput}
Vector of quaternionic Hermitian matrices with entries
                [1] 
 [1,] -1.746230e-10 
 [2,] -5.820766e-11 
 [3,] -1.746230e-10 
 [4,]  7.275958e-12 
 [5,]  2.910383e-11 
................... 
[41,]  4.365575e-11 
[42,]  0.000000e+00 
[43,]  4.365575e-11 
[44,] -9.458745e-11 
[45,] -1.455192e-10 
\end{Soutput}
\end{Schunk}

However, the Albert algebra does not satisfy the identity:

\begin{Schunk}
\begin{Sinput}
> G9(ralbert(1),ralbert(1),ralbert(1)) # Albert algebra does not obey G9:
\end{Sinput}
\begin{Soutput}
Vector of Albert matrices with entries
               [1]
    d1  32205.7845
    d2  37449.9386
    d3  -2734.0653
Re(o1)  21079.0754
 i(o1)  -8520.7870
 j(o1)   2463.1298
 k(o1)  11536.5988
 l(o1)  10866.9855
il(o1)  21487.3519
jl(o1)   -385.9625
kl(o1)  -7022.2586
Re(o2)  29565.4992
 i(o2) -10266.0920
 j(o2)   6242.0058
 k(o2)  15384.8470
 l(o2)   6507.9183
il(o2)   8053.8656
jl(o2)  -2563.1248
kl(o2)    304.0676
Re(o3)  20493.9088
 i(o3)  30960.4312
 j(o3) -12243.8005
 k(o3)  -6366.5065
 l(o3)   5962.5897
il(o3) -12700.1558
jl(o3) -24581.3730
kl(o3)   5110.3595
\end{Soutput}
\end{Schunk}

\bibliographystyle{apalike}
\bibliography{jordan_arxiv}

\end{document}